\documentstyle[epsf]{elsart}

\begin{document}
\begin{frontmatter}
\title{Axiomatic theory of nonequilibrium system}

\author{Alexander I. Olemskoi\thanksref{AO}}
\address{Sumy State University,
Riskier-Korsakov St. 2, 40007 Sumy, Ukraine}
\thanks[AO]{E-mail: olemskoi\char'100ssu.sumy.ua}

\begin{abstract}

Mutually conjugated synergetic schemes are assumed to address evolution of
nonequilibrium self-organizing system.
Within framework of the former, the system
is parameterized by a conserving order parameter being a density, a conjugate
field reducing to gradient of related flux, and control
parameter, whose driven magnitude fixes stationary state. We show
that so-introduced conjugate field and control parameter are relevant to
entropy and internal energy, so that self-organization effect is
appeared as a negative temperature. Along the line of the conjugated
scheme, roles of order parameter, conjugate field and control parameter are
played with a flux of conserving value, and gradients of both chemical
potential and temperature. With growth of the latter, relevant value of the
entropy shows to decrease in supercritical regime related to spontaneous
flux-state. We proof that both approach stated on using density and conjugated
flux as order parameters follow from unified field theory related to the
simplest choice of both Lagrangian and dissipative function.

\vspace{0.2cm}


\vspace{0.2cm}


\end{abstract}

\end{frontmatter}

\section{Introduction}

Nowadays, complex phenomena type of nonequilibrium phase transformations,
sy\-ner\-ge\-tics, self-organized criticality, etc. attract very much
attention \cite{1}, \cite{2}. That is stipulated not only applications to a
vast variety of concrete problems, among which are pattern formation,
morphgenesis, traffic flows at so one, but also due to lack of a general theory
of nonequilibrium systems, in contrast to equilibrium statisical physics based
on well-established concepts and tools (see \cite{2a}). 
The latter are known to be stated on
introduction of thermodynamic potentials and extremum principle \cite{3}. It is
quite clear that, within framework of the theory of nonequilibrium systems,
these ingredients must keep in relations respected to equilibrium limit, along
with Onsager-type relations between gradient of thermodynamic state
pa\-ra\-me\-ters and conjugated fluxes. This paper contains an attempt to build
up such a scheme originated from synergetic concept.

As is known, synergetics deals with far-from-equilibrium systems
in a variety of fields,
ranging from physics over chemistry and biology to sociology, where external
driven affects arrive at a self-organization \cite{4}. Formally, the latter is
appeared as entropy decrease with the subsystem moving off equilibrium --
in contrast to the second law of thermodynamics that is
applicable to whole system,
which keeps in equilibrium as rule. Being metastable in nature, such ordered
state tends to a local minimum of synergetic potential, where it may be located
during very long time intervals.
Main difference of such states is to be further stationary,
than equilibrium one.

In general case, two types of stable states are possible: (i) with
nonzeroth magnitude of an order parameter and (ii) with the same of a
conjugated flux. It is appeared that the only first state is essential
in the case when relevant order parameter is nonconserving. In this case, the
synergetic approach addresses kinetic of nonequilibrium phase transition.
However, in much more complicated case of a conserving order parameter
a flux state could appear spontaneously.
In this connection, a problem arise to treat this state
in analogy with synergetic theory of phase transition \cite{5}.
A method adopted in this work is stated on a generalization of such approach to
a general theory of nonequilibrium systems
based on axiomatic field theory.

The paper is organized in the following manner.
In Section 2, a synergetic scheme
based on the famous Lorenz system is assumed to address evolution of
self-organizing system parameterized by a conserving order parameter, a
conjugate field, reducing to gradient of related flux, and
a control parameter, whose externally driven magnitude fixes stationary state of
the system under consideration. We show, within adiabatic approximation, that
in the case of nonequilibrium thermodynamic system so-introduced conjugate
field and control parameter are relevant to the entropy and the internal
energy, respectively. As a result, self-organization effect is expressed in
appearance of a negative temperature, which magnitude decreases monotonically
with growth of driven control parameter. Section 3 is devoted to study of flux
state, whose parameterization is achieved by introducing, apart from a flux, of
gradients of both chemical potential and temperature to play role of the
conjugate field and the control parameter, respectively. With growth of the
latter, relevant value of the entropy shows to increase monotonically in
subcritical regime and decrease in supercritical one related to spontaneous
flux-state. Principle particularity of self-consistent equations used in
Sections 2 and 3 is that these are postulated to be reduced to the form of
familiar Lorenz scheme. The only justification of such a choice is that
relevant steady state relations lead to the standard equalities of the
equilibrium thermodynamics. Section 4 proofs that both approach
stated on using density and conjugated flux as order parameters are followed
from unified field theory for a set of pseudo-vectors composed of: (i)
density and flux; (ii) entropy and gradient of chemical potential;
(iii) internal energy and temperature gradient. It is appeared that
Lagrangian and dissipative function related to the Lorenz scheme are the
simplest ones: the former contains quadratic and cubic terms only, whereas the
latter -- quadratic time derivative contributions, whose prolonging accounts
dissipation processes.

\section{Density as order parameter}

We address a system with density $\rho=\rho({\bf r})$ and velocity ${\bf V}={\bf V}({\bf r})$
distributed over space according to the following decompositions:
\begin{eqnarray}
\rho\equiv\rho_0\left[1+\eta({\bf r})\right];\qquad
{\bf V}\equiv V_0\left[{\bf 1} + {\bf v}({\bf r})\right],\quad{\bf 1}\equiv{{\bf V}_0\over V_0}.
\label{1}
\end{eqnarray}
Here $\rho_0, {\bf V}_0={\rm const}$ are volume averaged values,
$\eta({\bf r})$, ${\bf v}({\bf r})$ are space-dependent dimensionless components.
Relevant flux and energy
\begin{eqnarray}
{\bf Q}\equiv\rho{\bf V},\quad
E\equiv{\rho\over 2}{\bf V}^2
\label{2}
\end{eqnarray}
are determined by these components as follows:
\begin{eqnarray}
&{\bf Q}\equiv Q_0[(1+\eta){\bf 1} + {\bf q}({\bf r})],\quad
E\equiv E_0[(1+\eta)+2q({\bf r})+\epsilon({\bf r})];&\nonumber\\
\label{3}\\
&Q_0\equiv\rho_0 V_0,\quad
E_0\equiv {Q_0^2\over 2\rho_0}.&\nonumber
\end{eqnarray}
Statistical state is fixed by internal components
\begin{eqnarray}
{\bf q}\equiv (1+\eta){\bf v},\quad
\epsilon\equiv{{\bf q}^2\over 1+\eta}.
\label{4}
\end{eqnarray}

A conserved value of density $\rho$ is governed by the continuity equation
\begin{eqnarray}
{\partial\rho\over\partial t}+\nabla {\bf Q}=0,\qquad \nabla\equiv{\partial\over\partial{\bf r}},
\label{5}
\end{eqnarray}
where ${\bf r}$, $t$ are coordinate and time.
With switching a dissipative sink $-\eta/\tau$ characterized with a
relaxation time $\tau$, the continuity equation (\ref{5}) takes the form
\begin{eqnarray}
{\partial\eta\over\partial t}+V_0\nabla {\bf q}=-{\eta\over\tau},\quad V_0\equiv{Q_0\over\rho_0}.
\label{7}
\end{eqnarray}
Introducing source term
\begin{eqnarray}
s\equiv -\lambda\nabla {\bf q},\quad\lambda\equiv V_0\tau,
\label{8}
\end{eqnarray}
one reduces this equation to the simplest form
\begin{equation}
\tau\dot\eta= -\eta + s,
\label{9}
\end{equation}
where dot stands for time derivative. At stationary state ($\dot\eta=0$) this
equation shows coincidence of the source $s$ and the density deviation $\eta$.

Let us find now an evolution equation for the source (\ref{8}) related to the
dimensionless flux ${\bf q}$. This equation should be obviously reduced to
Onsager relation
\begin{eqnarray}
{\bf q}=-\lambda\nabla\eta
\label{10}
\end{eqnarray}
at stationary state, where the source takes the form
\begin{eqnarray}
s=\lambda^2\nabla^2\eta.
\label{11}
\end{eqnarray}
On the other hand, here $s=\eta$, so that the dependence $\eta({\bf r})$
is governed by the Poisson equation
\begin{eqnarray}
\lambda^2\nabla^2\eta = \eta.
\label{12}
\end{eqnarray}
To be subjected to above conditions, we postulate the evolution equation found
in the following form:
\begin{eqnarray}
\tau_s\dot s= - s + \eta\epsilon.
\label{13}
\end{eqnarray}
Here, the first term in right-hand side is addressed relaxation process characterized with time $\tau_s$,
the second term is non-linear in the form due to a parameter $\epsilon$, being introduced
to take magnitude $\epsilon=1$ at stationary conditions.
Time evolution of this parameter is supposed to be of relaxational type also,
but if the autonomously evolving values $\eta(t)$, $s(t)$ go to zero with time $t\to\infty$,
the parameter $\epsilon(t)$ is supposed to tend to a finite magnitude $\epsilon_0\ne 0$,
within an autonomous regime.
We postulate the evolution equation for time-dependence $\epsilon(t)$
of such a parameter as follows:
\begin{eqnarray}
\tau_\epsilon\dot\epsilon=(\epsilon_0 - \epsilon) - \eta s,
\label{14}
\end{eqnarray}
where $\tau_\epsilon$ is relevant relaxation time.

Being equivivalent to the Lorenz form (see Section 4),
equations (\ref{9}), (\ref{13}),
(\ref{14}) obtained constitute the basis for self-consistent description of the
 evolving system with driven control parameter $\epsilon_0$ and relaxation times
$\tau_\eta$, $\tau_s$, $\tau_\epsilon$. The distinguishing feature of these
equations is that nonlinear terms, that enter Eqs.~(\ref{13}), (\ref{14}), are
of opposite signs, while Eq.~(\ref{9}) is linear. Physically, the latter means
that on the early stage of the process under consideration, the dimensionless
density deviation $\eta$ is reduced to the source $s$ defined Eq.~(\ref{8}).
The negative sign of the last term in Eq.~(\ref{14}) can be regarded as a
manifestation of Le Chatelier principle. Indeed, as we convince below,
supercritical $\epsilon$-increase results in self-organization process, whereas
the density deviation $\eta$ and the source $s$ in Eq.~(\ref{14}) tend to
suppress the growth of this parameter. The positive feedback of $\eta$ and
$\epsilon$ on $s$ in Eq.~(\ref{13}) plays a fundamental part in the problem. As
we shall see later, it is precisely the reason behind the self-organization
that brings about the self-organization.

In general case, the system (\ref{9}), (\ref{13}), (\ref{14}) cannot be
solved analytically, but in the simplest case, where $\tau_s,
\tau_\epsilon\ll\tau_\eta$, the source $s$ and the control parameter $\epsilon$
can be eliminated by making use of the adiabatic approximation that implies
neglecting of the left-hand sides of Eqs.~(\ref{13}), (\ref{14}). As a result,
the dependencies of $\epsilon$ and $s$ on $\eta$ are given by
\begin{equation}
s=\frac{\epsilon_{0}\eta}{1+\eta^{2}},\quad\epsilon=\frac{\epsilon_{0}}{1+\eta^{2}}.
\label{15}
\end{equation}
Note that, under $\eta$ is within the physically meaningful range between $0$
and $1$, the parameter $\epsilon$ is a monotonically decreasing function of the
density deviation $\eta$, whereas the source $s$ increases with $\eta$ (at
$\eta>1$ we have ${\rm d}s/{\rm d}\eta<0$ and the self-organization process
becomes unstable). Such type behaviour permit to assume that the source $s$ can
be meant as the entropy increase caused by the density deviation during
self-organization, whereas the parameter $\epsilon$ is a related energy. Within
the framework of such supposition, the dependence
\begin{equation}
s=\sqrt{\epsilon(\epsilon_{0}-\epsilon)}
\label{16}
\end{equation}
following from Eqs.~(\ref{15}), arrives at the temperature
\begin{equation}
T\equiv{\partial\epsilon\over\partial s}
\label{16.1}
\end{equation}
as follows:
\begin{eqnarray}
T=-\left(1-{\epsilon_{0}\over 2\epsilon}\right)^{-1}\sqrt{{\epsilon_{0}\over\epsilon} - 1}.
\label{17}
\end{eqnarray}
So defined temperature increases monotonically with energy growth from
magnitude $T=0$ at $\epsilon=0$ to infinity at the point
$\epsilon=\epsilon_{0}/2$. Here, the temperature $T$ breaks abruptly to
negative infinity and then increases monotonically again to initial magnitude
$T=0$ at $\epsilon=\epsilon_{0}$. This means that inside domain
$0\leq\epsilon<\epsilon_{0}/2$ the self-organization process is dissipative to
behave in usual manner; contrary, within domain
$\epsilon_{0}/2<\epsilon\leq\epsilon_{0}$ self-organization process evolves so,
that an energy increase derives to entropy decrease, in accordance with
negative value of temperature.

Substitution of the first Eq.~(\ref{15}) into Eq.~(\ref{9}) yields the
Landau-Khalatnikov equation
\begin{equation}
\dot{\eta}=-\frac{\partial W}{\partial\eta}
\label{18}
\end{equation}
with synergetic potential
\begin{equation}
W=\frac{1}{2}\eta^{2}-
\frac{\epsilon_{0}}{2}\ln{\left(1+\eta^{2}\right)}.
\label{19}
\end{equation}
For $\epsilon_{0}<1$, the $\eta$-dependence of $W$ is monotonically increasing
and the only stationary value of $\eta$ equals zero, $\eta_{e}=0$. If
externally driven energy $\epsilon_{0}$ exceeds the critical value,
$\epsilon_{c}=1$, the synergetic potential assumes the minimum with nonzero
steady state magnitudes of density deviation and entropy
$\eta_{e}=s_{e}=\sqrt{\epsilon_{0}-1}$ to be dependent on the driven energy
$\epsilon_{0}$ and constant value of the stationary energy $\epsilon_{e}=1$.
The temperature (\ref{17}) takes the stationary magnitude
\begin{eqnarray}
T_0=-~{\sqrt{\epsilon_0 - 1}\over 1-\epsilon_0/2}
\label{20}
\end{eqnarray}
being negative within supercritical domain $1\leq\epsilon_0<2$. Thus, the
stationary temperature $T_0$ decreases monotonically with the driven energy
increase from the zeroth magnitude at $\epsilon_0=1$ to negative infinity at
$\epsilon_0=2$. This means that self-organization stimulus increases with
growth of nonequilibrium power.

\section{Flux as order parameter}

The above consideration shows that the dissipative dynamics of the conserved
system can be represented within the framework of the Lorenz model, where the
density deviation $\eta$ plays a role of the order parameter, the source
(\ref{8}) is reduced to the entropy $s$ being conjugated field to the order
parameter and the internal energy $\epsilon$ is the control parameter. The key
point of this approach is that one postulates to address the density deviation
$\eta({\bf r}, t)$ as fundamental field. However, a variety of the physical
systems are known as well \cite{1} -- \cite{2a}, where the flux deviation ${\bf
q}$ takes a fundamental part of the order parameter. To study relevant systems,
we postulate the Lorenz system in the following form 
(cf. Eqs.~(\ref{9}), (\ref{13}), (\ref{14}))
\begin{equation}
\tau_{\bf q}\dot{\bf q}=-{\bf q} + {\bf f},
\label{9a}
\end{equation}
\begin{equation}
\tau_{\bf f}\dot{\bf f}=-{\bf f} + {\bf 1}_{\bf f}({\bf q}{\bf g}),
\label{13a}
\end{equation}
\begin{equation}
\tau_{\bf g}\dot{\bf g}=({\bf g}_0-{\bf g}) - {\bf 1}_{\bf g}({\bf q}{\bf f}).
\label{14a}
\end{equation}
Here, $\tau_{\bf q}$, $\tau_{\bf f}$, $\tau_{\bf g}$ are relaxation times,
${\bf 1}_{\bf f}$, ${\bf 1}_{\bf g}$ are unit vectors along related directions
and we introduce a constant vector ${\bf g}_0\equiv g_0{\bf 1}_0$, which physical meaning will be determined.
At the stationary state, when $\dot{\bf q}=0$, a force ${\bf f}$ reduces to the flux ${\bf q}$.
This means that in non-stationary case $\dot{\bf q}\ne 0$ this force
takes the form of generalized Onsager equality (cf. Eq.~(\ref{10}))
\begin{eqnarray}
{\bf f}=-\lambda\nabla\mu,
\label{10a}
\end{eqnarray}
where a specific thermodynamic potential $\mu$ is introduced to be the chemical
potential in usual case. Thus, we can conclude that the vector ${\bf f}$ is
reduced to the thermodynamic force.

To state a physical meaning of the third vector ${\bf g}$, let us suppose, as
above, that relaxation times are subjected to conditions $\tau_{\bf f},
\tau_{\bf g}\ll\tau_{\bf q}$. Then, within the adiabatic appro\-xi\-ma\-tion 
one obtains:
\begin{eqnarray}
{\bf f}=\frac{g_{0}({\bf 1}_0{\bf 1}_{\bf q}){\bf 1}_{\bf f}~q}
{1+({\bf 1}_{\bf g}{\bf 1}_{\bf q})({\bf 1}_{\bf f}{\bf 1}_{\bf q}){\bf q}^{2}},\quad
{\bf g}=\frac{{\bf 1}_0 + g_{0}({\bf 1}_{\bf g}{\bf 1}_{\bf f})[[{\bf 1}_{\bf g}{\bf 1}_0]{\bf 1}_{\bf q}]~
{\bf q}^{2}}
{1+({\bf 1}_{\bf g}{\bf 1}_{\bf q})({\bf 1}_{\bf f}{\bf 1}_{\bf q}){\bf q}^{2}},
\label{15a}
\end{eqnarray}
where parentheses and square brackets denote scalar and vector productions,
respectively. These equalities are reduced to form of the
relations ~(\ref{15}) if
the vector ${\bf g}$ do not varies its direction, i. e.,
$\left[{\bf 1}_{\bf g}{\bf 1}_0\right]=0$.
Moreover, it is enough for our aims to restrict ourselves addressing
the simplest one-dimensional case. Here, the expressions (\ref{15a}) give the
following state equation (cf. Eq. (\ref{16}))
\begin{equation}
f=\sqrt{g(g_{0}-g)}.
\label{16a}
\end{equation}
Then, a derivative
\begin{equation}
{\partial f\over\partial g}\equiv -s
\label{16.2}
\end{equation}
takes the form (cf. Eq. (\ref{17})):
\begin{eqnarray}
s=\left(1-{g_{0}\over 2g}\right)\left({g_{0}\over g} - 1\right)^{-{1\over 2}}.
\label{17a}
\end{eqnarray}
With $g$-growth within the domain $0<g<g_0$, this derivative increases
monotonically from $-\infty$ to $\infty$ taking the magnitude $s=0$ at
$g=g_0/2$. This means physically that the quantity defined by equality
(\ref{16.2}) could be recognized as a specific entropy infinitely increasing
with tending the control parameter $g_0/2\leq g<g_0$ to externally driven value
$g_0$. Respectively, the control parameter ${\bf g}$ gets meaning of the
temperature gradient taken with inverted sign:
\begin{equation}
{\bf g}\equiv -\lambda\nabla T.
\label{18a}
\end{equation}
At the stationary condition, when $\dot{\bf q}=0$, one obtains the steady-state
magnitudes of flux, thermodynamic force and temperature gradient as follows:
\begin{equation}
q_{e}=f_{e}=\sqrt{g_{0}-1},\quad g_{e}=1.
\label{19a}
\end{equation}
On the other hand, steady-state specific entropy (cf. Eq. (\ref{20}))
\begin{eqnarray}
s_0={1-g_0/2\over\sqrt{g_0 - 1}}
\label{20a}
\end{eqnarray}
decreases monotonically from infinity to zero with the temperature gradient
growth within supercritical domain $1<g_0<2$. Thus, in accordance with above
supposition, self-consistent evolution of flux ${\bf q}$, thermodynamic
force ${\bf f}$ and sign inverted temperature gradient ${\bf g}$ arrives at
entropy decrease that means self-organization process at supercritical
magnitudes of the control parameter $g_0>1$.

\section{Field theory of self-organizing system}

According to our previous contribution \cite{6}, the Lorenz system is
relevant microscopically to the bozon-fermion Hamiltonian of Dicke type.
Phenomenologically, its consideration is possible only within supersymmetry
field theory \cite{7}. To take into account quite different commutation rules
related to different freedom degrees, let us introduce some pseudo-vectors
\begin{eqnarray}
{\vec\phi} =\left(
\begin{array}{ll}
\eta\\ {\bf q}
\end{array}\right),~~
{\vec{\cal F}}=\left(
\begin{array}{ll}
s \\ {\bf f}
\end{array}\right),~~
{\vec{\cal C}}=\left(
\begin{array}{ll}
\varepsilon\\ {\bf\delta}
\end{array}\right);\quad
{\vec{\cal F}}{\vec{\cal C}}\equiv - {\vec{\cal C}}{\vec{\cal F}},~~
{\bf f}\equiv -\lambda\nabla\mu,
\label{22}
\end{eqnarray}
to yield material fields ${\vec\phi}$, as well as 
mutually anticommuting components of conjugate fields ${\vec{\cal F}}$ 
and state parameters ${\vec{\cal C}}$ 
(here, parameters $\delta$, $\varepsilon$ have to be defined). The system
behaviour is postulated to be governed by Lagrangian
\begin{eqnarray}
{\cal L}={\cal L}_0 + {\cal L}_{int}
\label{23}
\end{eqnarray}
that comprises of a bare part
\begin{eqnarray}
{\cal L}_0=
{\vec\phi}{\vec{\cal F}}{\vec{\cal C}}_0,\quad
{\vec{\cal C}}_0\equiv\left(
\begin{array}{ll}
\epsilon_0 \\ {\bf g}_0
\end{array}\right),
\label{23a}
\end{eqnarray}
and an interaction contribution
\begin{eqnarray}
{\cal L}_{int}=
-{\vec\phi}{\vec{\cal F}}{\vec{\cal C}}.
\label{23b}
\end{eqnarray}
In analogy, dissipative function
\begin{eqnarray}
{\cal R}={\cal R}_m + {\cal R}_f
\label{24}
\end{eqnarray}
consists of a material component
\begin{eqnarray}
{\cal R}_m=
{1\over 2}({\vec{\cal C}}_0{\vec{\cal D}}_{\phi}{\vec\phi})^2
\label{24a}
\end{eqnarray}
and field gradient terms
\begin{eqnarray}
{\cal R}_f = {1\over 2}({\vec{\cal D}}_{\cal F}{\vec{\cal F}})^2 +
{1\over 2}({\vec{\cal D}}_{\cal C}{\vec{\cal C}})^2,
\label{24aa}
\end{eqnarray}
where the notions are introduced as follows:
\begin{eqnarray}
{\vec{\cal D}}_z\equiv{\vec{\tau}}_z{\partial\over\partial t}+\vec{ 1};~~~~
{\vec{\tau}}_{\phi}\equiv\left(
\begin{array}{ll}
\tau \\ \tau_{\bf q}
\end{array}\right),~
{\vec{\tau}}_{\cal F}\equiv\left(
\begin{array}{ll}
 \tau_s \\ \tau_{\bf f}
\end{array}\right),~
{\vec{\tau}}_{\cal C}\equiv\left(
\begin{array}{ll}
\tau_\epsilon \\ \tau_{\bf g}
\end{array}\right),~~
\vec{1}\equiv\left(
\begin{array}{ll}
1\\ 1
\end{array}\right),
\label{24b}
\end{eqnarray}
which are determined by prolonging derivatives ${\vec{\cal D}}_z$ to take into
account Debay-type relaxation given by times ${\vec{\tau}}_z$ with
$z={\vec\phi}, {\vec{\cal C}}, {\vec{\cal F}}$. Then, Euler equations
\begin{equation}
{\partial{\cal L}\over\partial{\rm z}}-
{\cal D}_z{\partial{\cal L}\over\partial({\cal D}_z{\rm z})}=
{\partial{\cal R}\over\partial({\cal D}_z{\rm z})},\quad
{\rm z}\equiv\vec{\phi}, \vec{\cal F}, \vec{\cal C},
\label{25}
\end{equation}
arrive at basic system of differential equations for
pseudo-vectors (\ref{22}):
\begin{eqnarray}
{\vec{\cal C}}_0{\vec{\cal D}}_\phi{\vec{\phi}}&=&
{\vec{\cal F}}{\vec{\cal C}}_0 -
{\vec{\cal F}}{\vec{\cal C}},
\nonumber\\
{\vec{\cal D}}_{\cal F}{\vec{\cal F}}&=&
{\vec{\cal C}}_0 {\vec{\phi}} -
{\vec{\phi}}{\vec{\cal C}},
\label{26}\\
{\vec{\cal D}}_{\cal C}{\vec{\cal C}}&=&
{\vec{\phi}}{\vec{\cal F}},
\nonumber
\end{eqnarray}
where the second equality takes into account anticommutation relation
for pseudo-vectors ${\vec{\cal F}}$ and ${\vec{\cal C}}$.
Equations (\ref{26}) are reduced to the form
\begin{eqnarray}
\tau\dot\eta&=&-\eta+(1-\varepsilon/\epsilon_0)s,
\nonumber\\
\tau_s\dot s&=&-s+\epsilon_0\eta-\eta\varepsilon,
\label{34}\\
\tau_\epsilon\dot\varepsilon&=&-\varepsilon+\eta s;
\nonumber
\end{eqnarray}
\begin{eqnarray}
\tau_{\bf q}\dot{\bf q}&=&-{\bf q}+(1-\delta/g_0){\bf f},
\nonumber\\
\tau_{\bf f}\dot{\bf f}&=&
-{\bf f}+{\bf 1}_{\bf f}({\bf g}_0{\bf q})-{\bf 1}_{\bf f}({\bf\delta}{\bf q}),
\label{37}\\
\tau_{\bf g}\dot{\bf \delta}&=&
-{\bf\delta}+{\bf 1}_{\bf\delta}({\bf q}{\bf f})
\nonumber
\end{eqnarray}
that takes the standard appearance \cite{4}
if one takes into account determinations (\ref{22}) and
keeps the linear terms only in equations for $\dot\eta$ and $\dot{\bf q}$.
To reduce these systems to the form of equations
(\ref{9}), (\ref{13}), (\ref{14}) and
(\ref{9a}) -- (\ref{14a}), respectively,
one needs moreover to account relations
\begin{eqnarray}
\varepsilon\equiv\epsilon_0-\epsilon,\quad
\delta\equiv{\bf g}_0-{\bf g};\qquad
{\bf g}\equiv-\lambda\nabla T.
\label{32}
\end{eqnarray}

Formally, the systems (\ref{34}), (\ref{37}) differ from corresponding
equations (\ref{9}), (\ref{13}), (\ref{14}) and
(\ref{9a}) -- (\ref{14a}) to arrive at the synergetic potential
\begin{equation}
W=\frac{1}{2}\eta^{2}+
\frac{\epsilon_{0}}{2}\frac{1}{1+\eta^{2}}
\label{19aa}
\end{equation}
instead of the dependence (\ref{19}). At supercritical driven energy
$\epsilon_{0}>1$ this dependence takes the minimum at the point
$\eta_e^2=\sqrt{\epsilon_{0}}-1$ related to stationary synergetic potential
$W(\eta_e)=\sqrt{\epsilon_{0}}-1/2$, whose value is lower than initial
magnitude $W(0)=\epsilon_{0}/2$. However, one keeps in mind that, within a
phenomenological ideology, the expressions obtained have an asymptotic
character to coincide only in vicinity of the critical magnitude
$\epsilon_{0}=1$. As a result, we may use the simplest of approaches
suppressing nonlinear terms in the first equations of the systems (\ref{34}),
(\ref{37}).

\section*{Acknowledgments}\label{sec:level1}

Author is grateful to Professor A.A. Katsnelson for collaboration and
illuminating conversations, which promote this activity.

\vspace{0.4cm}

\begin{center}
{\bf FIGURE CAPTIONS}
\end{center}

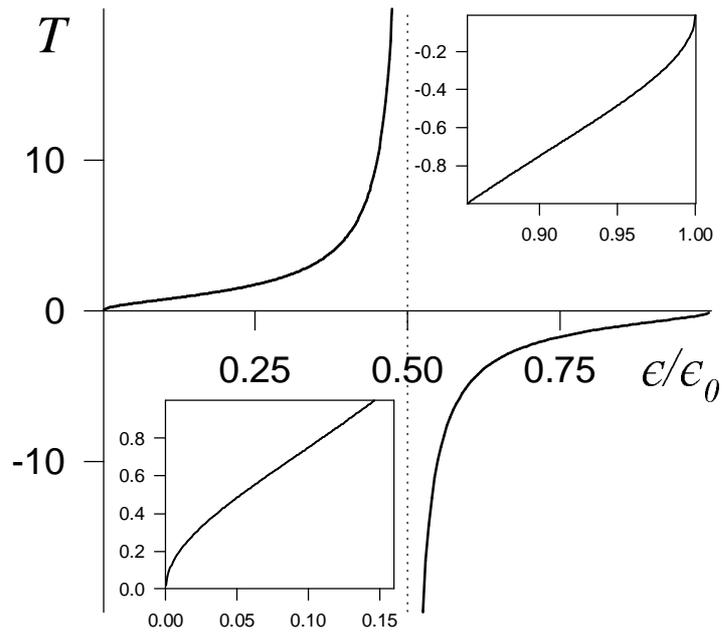
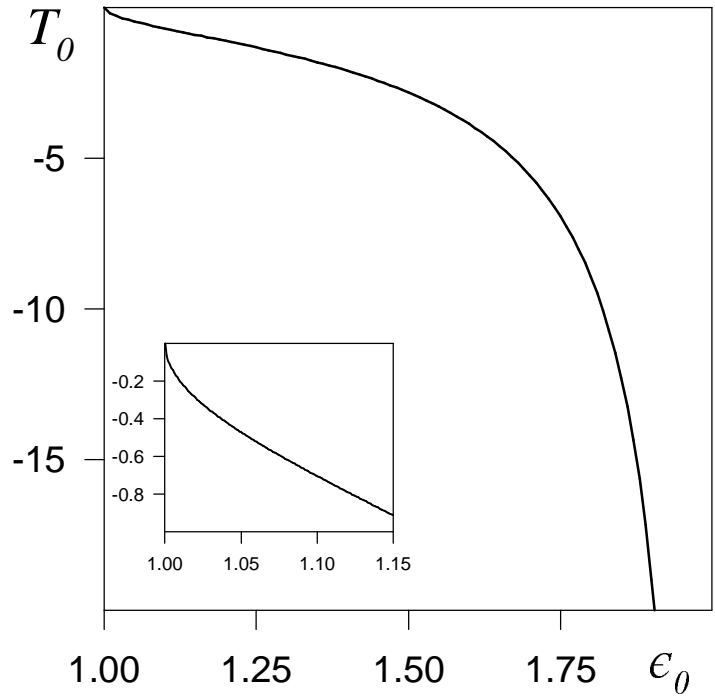
Fig.~1. The energy dependencies of the system temperatures: (a) nonstationary
magnitude $T$ versus ratio $\epsilon/\epsilon_0$; (b) stationary temperature
$T_0$ versus $\epsilon_0$.

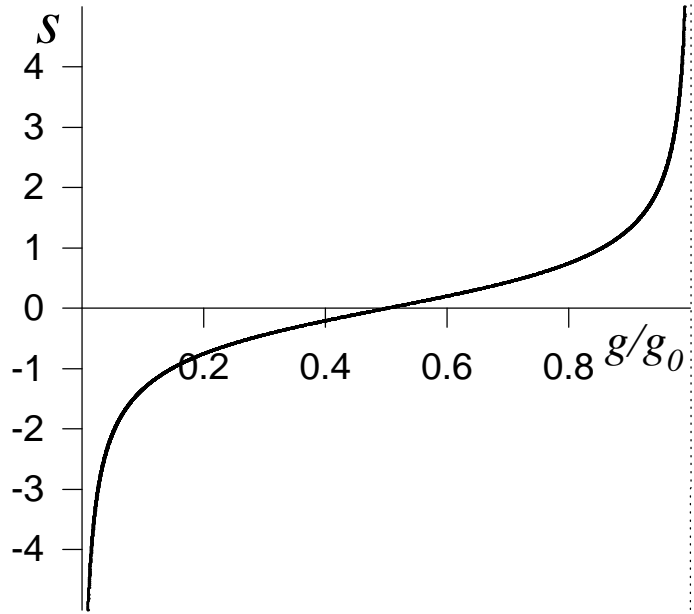
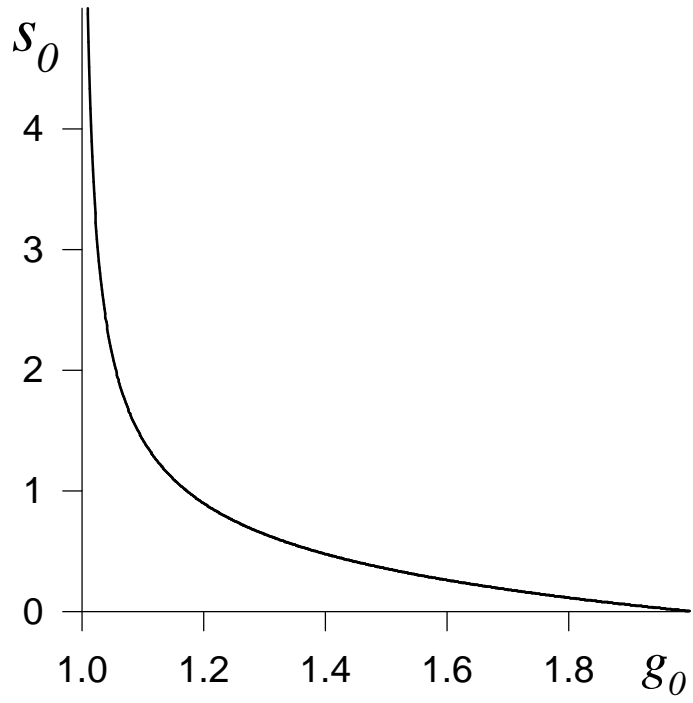
Fig.~2. The entropy dependencies on the temperature gradient: (a)
nonstationary magnitude $s$ versus ratio $g/g_0$; (b) stationary entropy
$s_0$ versus $g_0$.

\end{document}